\documentclass[aps,prb,twocolumn,letterpaper]{revtex4}

\usepackage{amsmath,amssymb,amsfonts,amsthm}
\usepackage{graphicx,bm}

\begin{document}

\title{Structural distortions and model Hamiltonian parameters: from
LSDA to a tight-binding description of LaMnO$_3$}

\date{\today}

\author{Claude Ederer}
\email{ederer@phys.columbia.edu}
\affiliation{Department of Physics, Columbia University, 538 West 120th
  Street, New York, NY 10027, U.S.A.} 
\author{Chungwei Lin}
\affiliation{Department of Physics, Columbia University, 538 West 120th
  Street, New York, NY 10027, U.S.A.}
\author{Andrew J. Millis}
\affiliation{Department of Physics, Columbia University, 538 West 120th
  Street, New York, NY 10027, U.S.A.}
 
\begin{abstract}
  The physics of manganites is often described within an effective
  two-band tight-binding (TB) model for the Mn $e_g$ electrons, which
  apart from the kinetic energy includes also a local ``Hund's rule''
  coupling to the $t_{2g}$ core spin and a local coupling to the
  Jahn-Teller (JT) distortion of the oxygen octahedra. We test the
  validity of this model by comparing the energy dispersion calculated
  for the TB model with the full Kohn-Sham band-structure calculated
  within the local spin-density approximation (LSDA) to density
  functional theory. We analyze the effect of magnetic order, JT
  distortions, and ``GdFeO$_3$-type'' tilt-rotations of the oxygen
  octahedra. We show that the hopping amplitudes are independent of
  magnetic order and JT distortions, and that both effects can be
  described with a consistent set of model parameters if hopping
  between both nearest and next-nearest neighbors is taken into
  account. We determine a full set of model parameters from the
  density functional theory calculations, and we show that both JT
  distortions and Hund's rule coupling are required to obtain an
  insulating ground state within LSDA. Furthermore, our calculations
  show that the ``GdFeO$_3$-type'' rotations of the oxygen octahedra
  lead to a substantial reduction of the hopping amplitudes but to no
  significant deviation from the simple TB model.
\end{abstract}

\pacs{}

\maketitle

\section{Introduction}
\label{sec:intro}

Manganite systems, $R_{1-x}A_x$MnO$_3$, where $R$ is a trivalent rare
earth cation (e.g. La$^{3+}$, Pr$^{3+}$, Nd$^{3+}$, \dots) and $A$ is
a divalent alkaline earth cation (e.g. Sr$^{2+}$, Ca$^{2+}$, \dots),
have attracted the attention of scientists already for
decades.\cite{Jonker/VanSanten:1950,Wollan/Koehler:1955,Coey/Viret/Molnar:1999,Tokura:2000,Dagotto/Hotta/Moreo:2001}
These compounds exhibit a very rich phase diagram as a function of
both temperature and composition, with various types of eventually
coexisting charge, orbital, and magnetic order, and they are therefore
important prototype materials to test our current understanding of
correlated electron systems. In addition, the observation of
``colossal magneto-resistance'',\cite{Jin_et_al:1994} a magnetic-field
induced change in electric resistivity by several orders of magnitude,
has spawned further interest both in the fundamental physics behind
this effect as well as in the question of whether this effect can be
utilized for technological applications.

\begin{figure}
\includegraphics[width=\columnwidth]{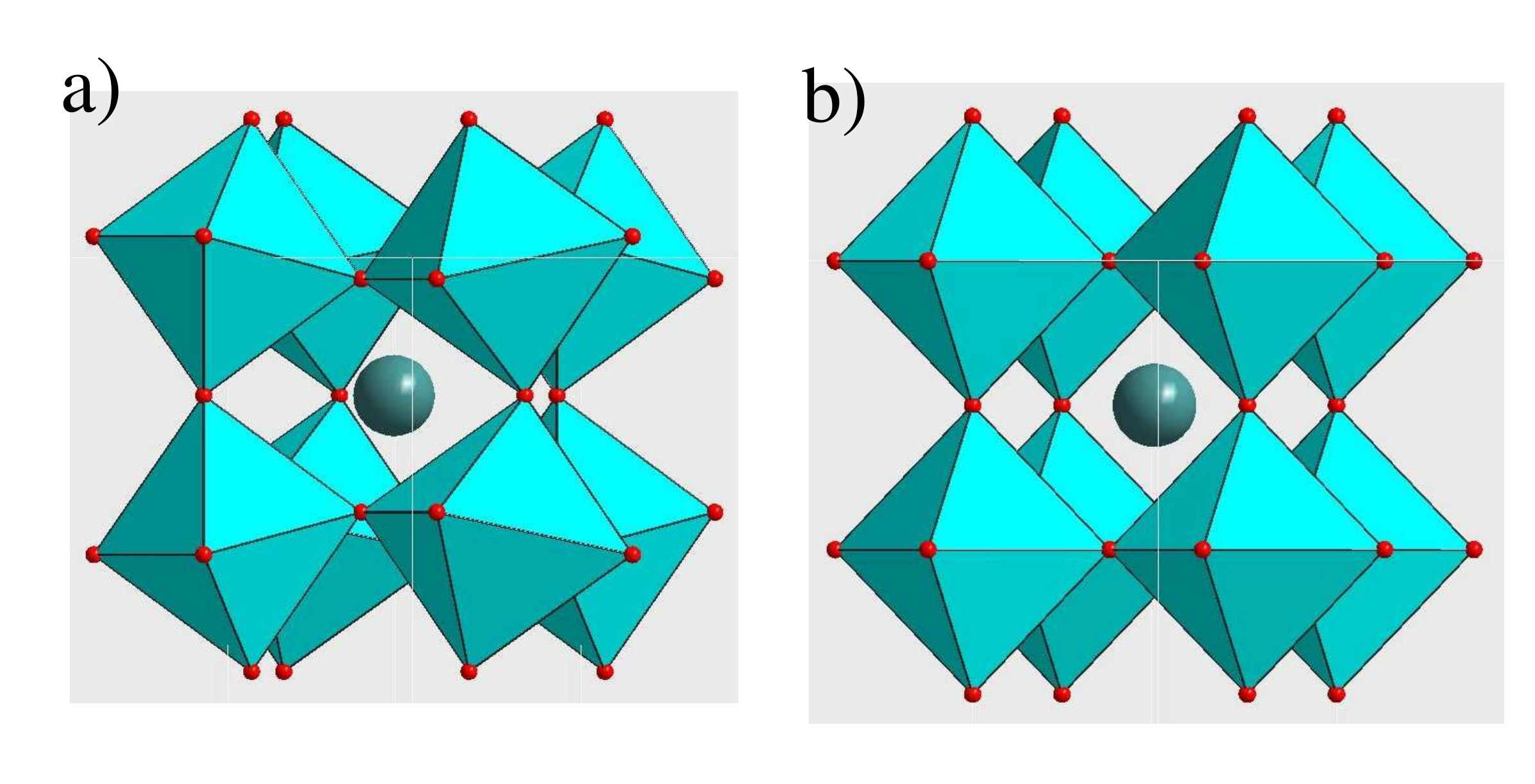}
\caption{a) Experimentally observed $Pnma$ structure of LaMnO$_3$
  according to Ref.~\onlinecite{Norby_et_al:1995}. b) Ideal cubic
  perovskite structure. The oxygen anions form a network of
  corner-shared octahedra. The Mn cations (not shown) are situated in
  the centers of the oxygen octahedra and the La cations occupy the
  space between the octahedra.}
\label{structure}
\end{figure}

LaMnO$_3$, the parent material for many manganite systems, exhibits an
orthorhombically distorted perovskite structure with $Pnma$ space
group (see Fig.~\ref{structure}a).\cite{Elemans_et_al:1971} The
observed deviation from the ideal cubic perovskite structure (shown in
Fig.~\ref{structure}b) involves both Jahn-Teller (JT) distortions of
the oxygen octahedra surrounding the Mn cations,\cite{Kanamori:1960}
as well as a collective tilting of these octahedra, the so called
``GdFeO$_3$-type'' distortion.\cite{Woodward:1997} The magnetic
moments of the Mn cations in LaMnO$_3$ order at $T_\text{N} = 140$\,K
in a so-called ``A-type'' antiferromagnetic
structure,\cite{Wollan/Koehler:1955} with parallel alignment of all
moments within a certain (001) plane and antiparallel alignment of the
moments between adjacent planes.

In the ideal cubic perovskite structure (see Fig.~\ref{structure}b)
the 3$d$ states of the Mn cations are split by the crystal field into
the lower-lying three-fold degenerate $t_{2g}$ states and the
higher-lying two-fold degenerate $e_g$ states. In LaMnO$_3$ the
majority-spin $t_{2g}$ states are fully occupied, whereas the two
majority-spin $e_g$ levels are filled with only one electron,
according to the formal high-spin $d^4$ electron configuration of the
Mn$^{3+}$ cation.

The theoretical modeling of manganite systems is usually based on the
assumption that the important low energy dynamics of these systems can
be described within an effective two band tight-binding (TB) model for
the Mn $e_g$ electrons. In this model, electrons can hop between the
$e_g$ orbitals at neighboring Mn sites, and the corresponding kinetic
energy term in the Hamiltonian is eventually complemented by local
terms describing the coupling to the $t_{2g}$ ``core spin'', the
coupling to the JT distortion of the surrounding oxygen octahedron,
and the electron-electron interaction. These models can account for
many of the properties observed in manganite systems (see
Ref.~\onlinecite{Dagotto/Hotta/Moreo:2001} for a recent
review). Reasonable values for the model parameters, which describe
the strengths of the various competing interactions, can be inferred
from experiments. Nevertheless, it is very desirable to independently
calculate values for these parameters based on fundamental ``first
principles'' theory. Evaluating the models using such independently
determined parameters provides a stringent test for the accuracy of
the model description.

Electronic structure calculations based on density functional theory
(DFT) \cite{Hohenberg/Kohn:1964,Kohn/Sham:1965} provide a way to study
the ground state electronic structure of a specific material without
having to resort to model assumptions, and therefore provide a good
starting point for the derivation of more simplified models and their
parameters (see e.g.  Refs.~\onlinecite{Gunnarsson_et_al:1989} and
\onlinecite{Hybertsen/Schlueter/Christensen:1989}). The electronic
structure of LaMnO$_3$ has been studied previously within the local
spin density approximation (LSDA) to DFT and by using the LSDA+$U$
method.\cite{Pickett/Singh:1996,Satpathy/Popovic/Vukajlovic:1996,Solovyev/Hamada/Terakura_PRL:1996,Terakura/Solovyev/Sawada:2000}
It was shown that many properties such as the correct magnetic ground
state and even some spectral properties are well described by these
methods, provided the correct experimental crystal structure is used
in the calculation.

Although the model treatment of manganite systems usually employs a
pure $e_g$ electron description, it is generally understood that the
electron hopping between the $e_g$ states on neighboring Mn sites is
truly an \emph{effective} hopping which is mediated by the
intermediate oxygen anions via $d$-$p$ or $d$-$s$ hopping. The
resulting bands with predominant $e_g$ character can be described by
an effective two-band model if the Mn $e_g$ states are energetically
separated from the oxygen $p$ and $s$ states. In this case, the
effective nearest neighbor hopping amplitude $t$ between the Mn $e_g$
states is (to leading order) given by:
\begin{equation}
\label{t_eff}
t \propto \frac{t_{pd}^2}{E_d - E_p} \quad .
\end{equation}
Here, $t_{pd}$ is the hopping amplitude between the Mn $e_g$ and the
oxygen $p$ states, $E_d$ and $E_p$ are the energies of the
corresponding ionic levels, and for simplicity we have neglected
hopping via the oxygen $s$ states.

The JT distortion changes the Mn-O bond lengths while the octahedral
tilts change the bond angles; thus both distortions affect the overlap
integrals which determine the hopping amplitude $t_{pd}$. It is
therefore not clear {\it a priori} that a simple effective TB model
with fixed (distortion-independent) hopping amplitudes $t$ can be used
to study the effects of lattice distortions in manganite systems.

Here, we use the Kohn-Sham band-structure calculated within the LSDA
as a reference for the non-interacting TB model, and we analyze how
well the relevant part of the energy dispersion of LaMnO$_3$ can be
fitted within an effective two-band TB model for the $e_g$
electrons. In particular, we analyze the effects of the two dominant
structural distortions in LaMnO$_3$, the JT distortion and the
GdFeO$_3$-type rotations, and we address the question of whether
magnetic and orbital (JT) order affects the effective hopping
amplitudes.

The result of our analysis is that the effective two-band model gives
a good fit of the $e_g$-projected Kohn-Sham band-structure, provided
that hopping between both nearest and next-nearest neighbors is taken
into account. We show that the same hopping amplitudes can be used for
the ferromagnetic, the A-type antiferromagnetic, and the JT distorted
case, so that the simple two-band TB model can be used to study the
effects of JT distortions.  Furthermore we quantify the dependence of
the hopping amplitudes on volume changes and on GdFeO$_3$-type
rotations. The latter lead to significant reductions of the hopping
amplitudes ($\sim$ 25\,\% for the experimental structure) relative to
the ideal cubic structure with the same unit cell volume. The hopping
amplitudes corresponding to the observed bond angles should therefore
be used in theoretical modeling.

Our results also provide a quantitative determination of the JT and
Hund's rule couplings. The result for the Hund's coupling is
consistent with previous work; the JT coupling is considerably smaller
than previous
estimates.\cite{Ahn/Millis:2000,Popovic/Satpathy:2000,Yin/Volja/Ku:2006}
We find that both the JT and Hund's coupling are required to stabilize
the insulating state within LSDA.

Our conclusions rely in an essential way on the energy separation of
the transition metal $d$-bands and the oxygen $p$-bands; methods such
as LSDA+$U$ which shift the energy of the transition-metal $d$-bands
relative to the energy of the oxygen $p$-bands can produce a band
structure that is very poorly described by a simple two-band TB model.

The remaining part of this paper is organized as
follows. Sec.~\ref{sec:methods} contains a brief summary of the
methods and technical details of our work. We first describe the
method we use for our LSDA calculations, then specify the TB
Hamiltonian, and finally describe how we decompose the various
structural distortions found experimentally in LaMnO$_3$. Our results
are discussed in Sec.~\ref{sec:results}, and we end with a summary of
our main conclusions and implications for future work.

\section{Methods and technical details}
\label{sec:methods}

\subsection{Computational method}

We calculate the LSDA Kohn-Sham band-structure for LaMnO$_3$ with both
ferromagnetic and A-type antiferromagnetic order in various structural
modifications using the projector augmented-wave (PAW) method
implemented in the ``Vienna Ab-initio Simulation Package''
(VASP).\cite{Bloechl:1994,Kresse/Furthmueller_PRB:1996,Kresse/Joubert:1999}
We treat the La 5$s$, La 5$p$, and Mn 3$p$ pseudo-core states as
valence states, and we use a plane-wave energy cutoff of 400\,eV in
all our calculations. We employ $\Gamma$-centered 6$\times$6$\times$6
and 4$\times$4$\times$3 $k$-point grids for the calculations
corresponding to the simple and quadrupled perovskite unit cells,
respectively, and corresponding grids for the structures in which the
unit cells are doubled along the $z$ direction or within the $x$-$y$
plane. These values result in a good convergence of the calculated
band-structures.

In order to extract the bands resulting from the Mn $e_g$ states we
use the ``fatbands'' technique, i.e. we assign a weight to each
calculated eigenvalue, which is proportional to the amount of Mn $e_g$
character contained in the corresponding Bloch function, and we
identify the $e_g$-derived bands as those containing a non-negligible
$e_g$ character.

It has been shown in Ref.~\onlinecite{Pickett/Singh:1996} that the
LSDA gives a good overall account of the electronic properties of
manganite systems, even though the tendency to open up an energy gap
between occupied and unoccupied states is underestimated within the
LSDA. This is a well-known feature of the LSDA, which results from the
inability of the LSDA to correctly account for the strong Coulomb
correlations between the rather localized $d$ states in transition
metal oxides. Such local Coulomb interactions are usually incorporated
in the model Hamiltonian via a separate interaction term. In the
following we do not include such an interaction term in our model
analysis, and thus the corresponding deficiencies of the LSDA do not
affect our results (assuming that the separate treatment of local
correlations is justified), except for the question related to the
energy separation between the Mn $d$ and the oxygen $p$ states, which
is discussed in Sec.~\ref{sec:summary}.

\subsection{Model Hamiltonian}

\begin{figure}
\includegraphics[width=0.49\columnwidth]{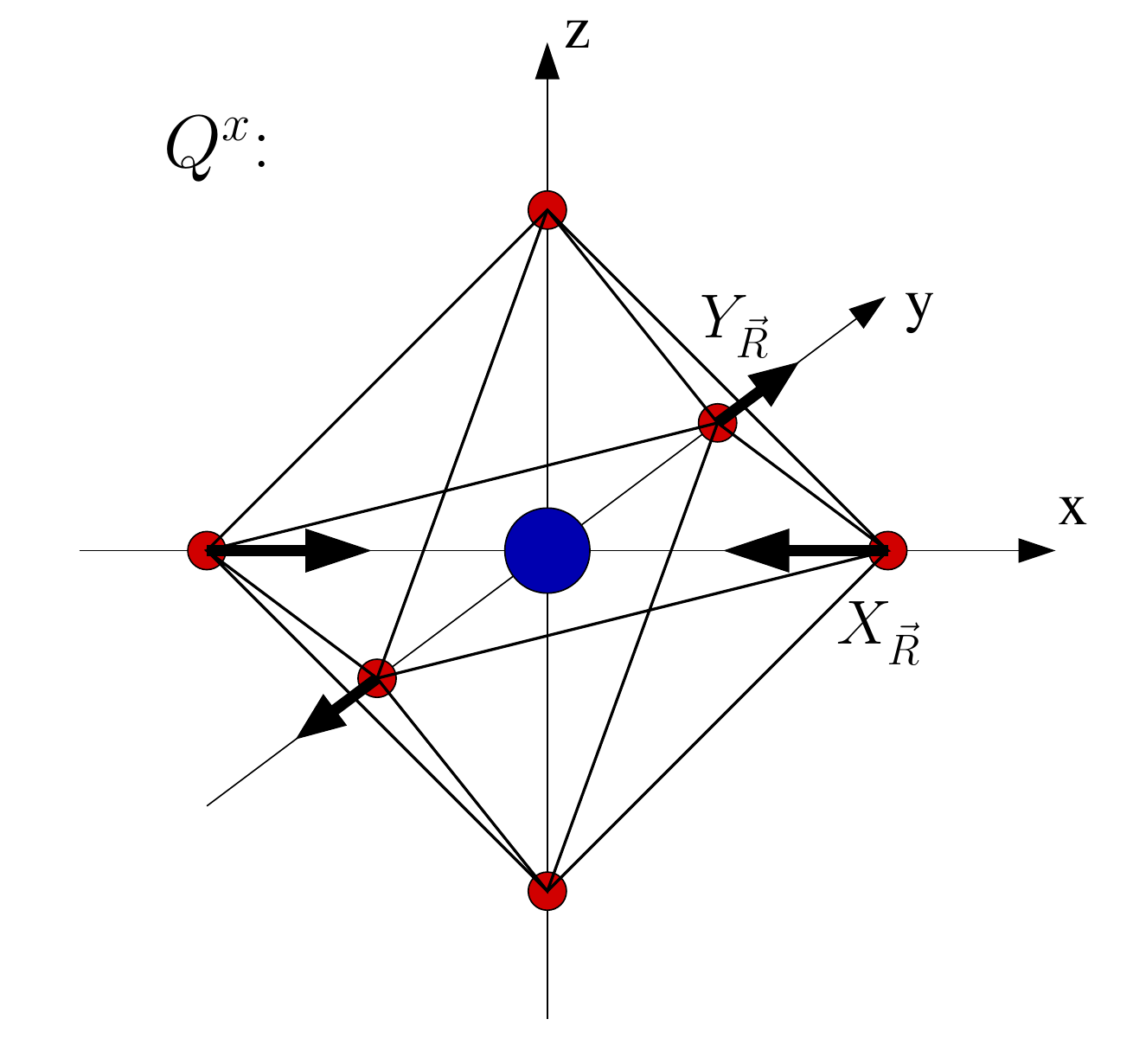}
\includegraphics[width=0.49\columnwidth]{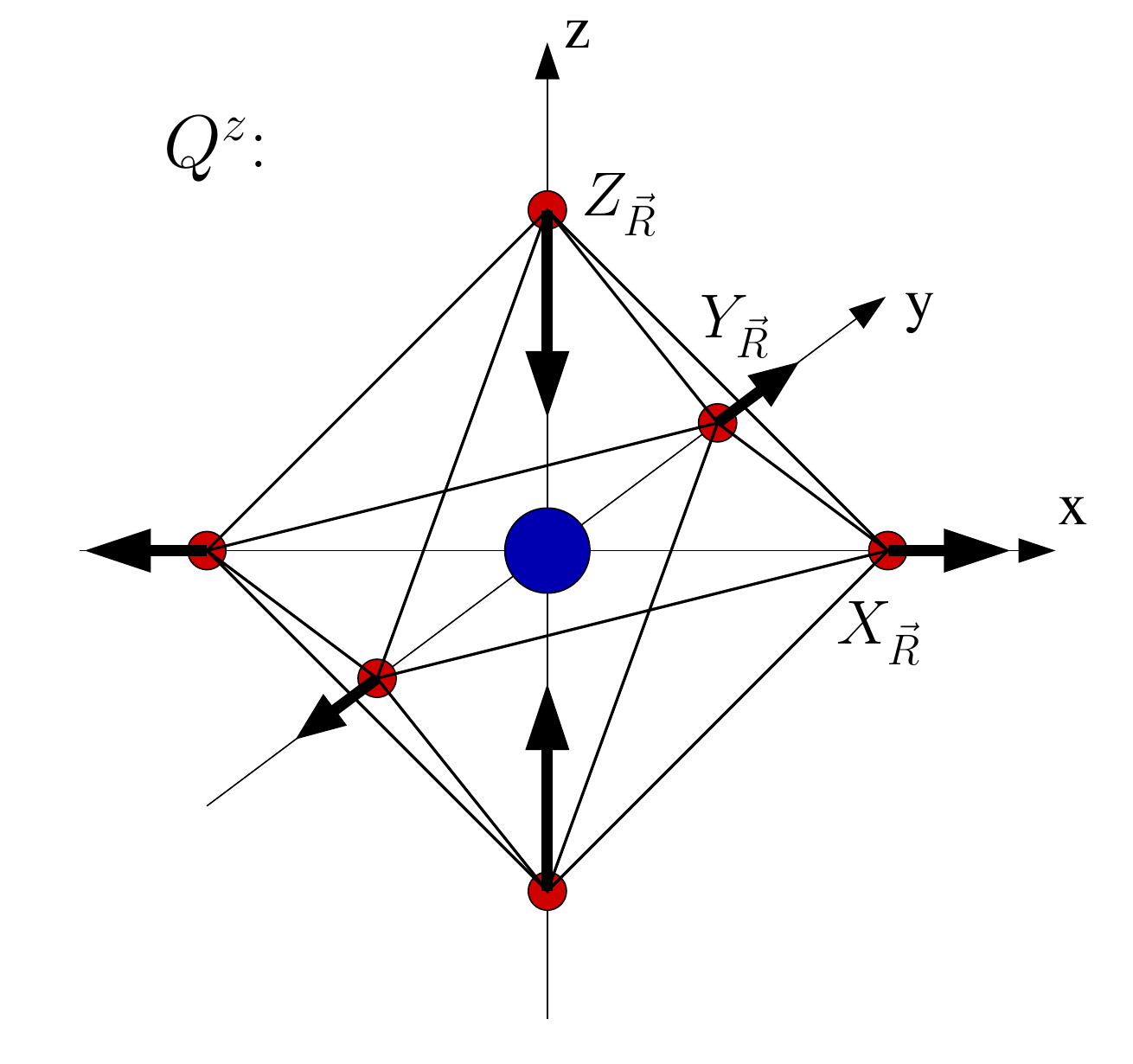}
\caption{Schematic depiction of the JT modes $Q^x$ (left side) and
  $Q^z$ (right side). The displacements $X_{\vec{R}}$, $Y_{\vec{R}}$,
  and $Z_{\vec{R}}$ used in Eqs.~(\ref{JT_amp}) and (\ref{JT_amp2})
  are indicated. Since we are considering only inversion symmetric
  distortions of the oxygen octahedra it is enough to specify the
  displacements of the oxygen anions adjacent to the central Mn cation
  in the positive $x$, $y$, and $z$ directions. Note that
  $|X_{\vec{R}}| = |Y_{\vec{R}}|$ in the case of $Q^x$ and
  $|X_{\vec{R}}| = |Y_{\vec{R}}| = \frac{1}{2}|Z_{\vec{R}}|$ for
  $Q^z$.}
\label{modes}
\end{figure}

In Sec.~\ref{sec:results} we relate the calculated LSDA band-structure
to the following TB model, which contains the terms that are typically
used for the theoretical modeling of manganite systems (see
e.g. Ref.~\onlinecite{Dagotto/Hotta/Moreo:2001}):
\begin{equation}
\hat{H} = \hat{H}_\text{kin} + \hat{H}_\text{Hund} + \hat{H}_\text{JT}
\quad ,
\label{model}
\end{equation}
with
\begin{align}
\label{ekin}
& \hat{H}_\text{kin} = - \sum_{\vec{R},\vec{\delta},\sigma} {\bm
    d}^+_{\vec{R},\sigma}  {\bm t}_{\vec{R},\vec{R}+\vec{\delta}} {\bm
    d}_{\vec{R}+\vec{\delta},\sigma} \quad ,  
\\
\label{hund}
& \hat{H}_\text{Hund} = - J \sum_{\vec{R},a,\sigma,\sigma'}
\vec{S}_{\vec{R}} \cdot \vec{\tau}_{\sigma,\sigma'} \, d^{+}_{\vec{R},a,\sigma}
    d_{\vec{R},a,\sigma'} \quad , 
\\ 
\label{jt}
& \hat{H}_\text{JT} = - \lambda \sum_{\vec{R},\sigma} \left(
    Q^x_{\vec{R}} \, {\bm d}^{+}_{\vec{R},\sigma} {\bm \tau}^x {\bm
      d}_{\vec{R},\sigma}  +
    Q^z_{\vec{R}} \, {\bm d}^{+}_{\vec{R},\sigma} {\bm \tau}^z {\bm
      d}_{\vec{R},\sigma} \right) \quad .
\end{align}
Here, $d_{\vec{R},a,\sigma}$ is the annihilation operator for an $e_g$
electron at site $\vec{R}$ in orbital $a$ with spin $\sigma$, and the
corresponding boldface symbol indicates a pseudo-spinor in orbital
space $\bm d_{\vec{R},\sigma} = (d_{\vec{R},1,\sigma},
d_{\vec{R},2,\sigma})^T$. The orbital indexes 1 and 2 correspond to
$\vert 3z^2-r^2 \rangle$ and $\vert x^2-y^2 \rangle$ orbitals,
respectively. ${\bm t}_{\vec{R},\vec{R}+\vec{\delta}} = \sum_{i=0}^{3}
t^i_{\vec{R},\vec{R}+\vec{\delta}} {\bm \tau}^{i}$ are the hopping
amplitudes between site $\vec{R}$ and $\vec{R}+\vec{\delta}$ and ${\bm
\tau}^i$ are the usual Pauli matrices supplemented by the 2$\times$2
unit matrix.  $\vec{S}_{\vec{R}}$ is the normalized core spin of the
$t_{2g}$ electrons ($|\vec{S}_{\vec{R}}| = 1$), and
$Q^{x,z}_{\vec{R}}$ are the amplitudes of the two JT modes at site
$\vec{R}$ that couple to the $e_g$ electrons:
\begin{align}
\label{JT_amp}
Q^x_{\vec{R}} & = \frac{1}{\sqrt{2}} \left( X_{\vec{R}} - Y_{\vec{R}}
\right) \quad , \\ Q^z_{\vec{R}} & = \frac{1}{\sqrt{6}} \left(
2Z_{\vec{R}} - X_{\vec{R}} - Y_{\vec{R}} \right) \quad .
\label{JT_amp2}
\end{align}
Here, $X_{\vec{R}}$, $Y_{\vec{R}}$, and $Z_{\vec{R}}$ are the
displacements along $\hat{x}$, $\hat{y}$, and $\hat{z}$ of the oxygen
anions that are situated adjacent to the Mn site at $\vec{R}$ in $x$,
$y$, and $z$ direction, respectively, and only inversion symmetric
distortions of the oxygen octahedra are taken into account (see
Fig.~\ref{modes}). $J$ and $\lambda$ are coupling constants for the
local interaction terms.

The first term in Eq.~(\ref{model}) describes the hopping between
neighboring Mn sites. We will consider hopping between both nearest
and next nearest neighbors. Symmetry dictates that the hopping
matrices for nearest neighbor hopping are:
\begin{align}
\label{hopping_start}
& {\bm t}_{\vec{R},\vec{R} \pm a\hat{x}} = \frac{t}{4} 
\begin{pmatrix} 1 & -\sqrt{3} \\ -\sqrt{3} & 3 \end{pmatrix} \\
& {\bm t}_{\vec{R},\vec{R} \pm a\hat{y}} = \frac{t}{4} 
\begin{pmatrix} 1 & \sqrt{3} \\ \sqrt{3} & 3 \end{pmatrix} \\
& {\bm t}_{\vec{R},\vec{R} \pm a\hat{z}} = t
\begin{pmatrix} 1 & 0 \\ 0 & 0 \end{pmatrix} \quad ,
\end{align}
and for next nearest neighbor hopping:
\begin{align}
& {\bm t}_{\vec{R},\vec{R} \pm a\hat{x} \pm a\hat{z}} = \frac{t'}{2} 
\begin{pmatrix} -2 & \sqrt{3} \\ \sqrt{3} & 0 \end{pmatrix} \\
& {\bm t}_{\vec{R},\vec{R} \pm a\hat{y} \pm a\hat{z}} = \frac{t'}{2} 
\begin{pmatrix} -2 & -\sqrt{3} \\ -\sqrt{3} & 0 \end{pmatrix} \\
& {\bm t}_{\vec{R},\vec{R} \pm a\hat{x} \pm a\hat{y}} = \frac{t'}{2}
\begin{pmatrix} 1 & 0 \\ 0 & -3 \end{pmatrix} \quad .
\label{hopping_end}
\end{align}
Here, $a$ is the lattice constant of the underlying cubic perovskite
lattice.

The second and third terms in Eq.~(\ref{model}) describe the Hund's
rule coupling to the $t_{2g}$ core spin and the coupling to the JT
distortion of the oxygen octahedra surrounding site $\vec{R}$,
respectively. The normalized $t_{2g}$ core spin is treated as a
classical vector which is fixed externally in our model
calculations. The values of $Q^{x/z}_{\vec{R}}$ are given by the
positions of the oxygen anions used in our LSDA calculations and are
also treated as external parameters in the TB model.

\subsection{Structural decomposition}
\label{sec:struc}

\begin{figure}
\includegraphics[width=\columnwidth]{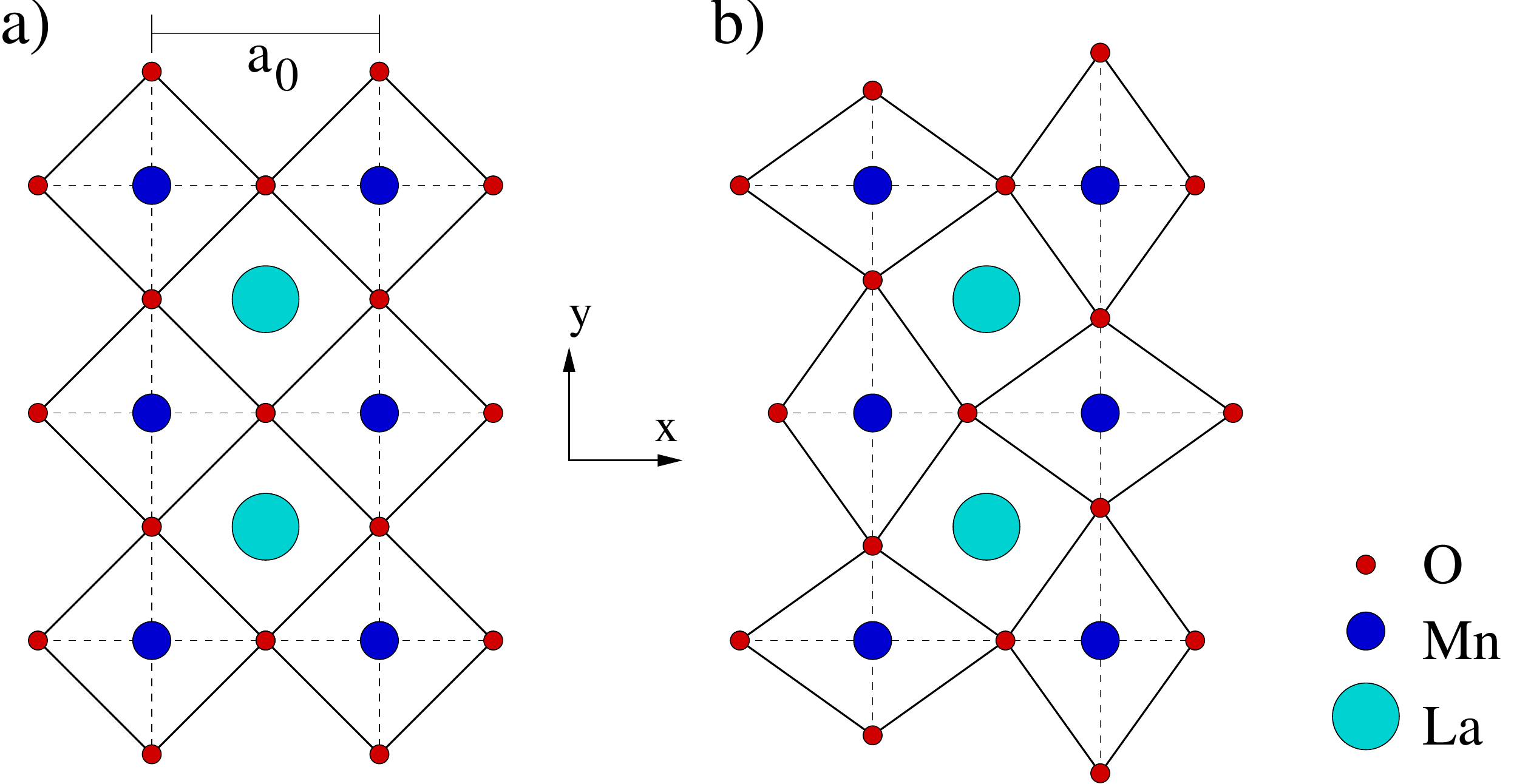}
\caption{Staggered $Q^x$-type JT distortion within the $x$-$y$ plane
found experimentally in LaMnO$_3$. The long and short axes of the
octahedra alternate along the $x$ and $y$ directions, as shown in b),
compared to the ideal structure shown in a).}
\label{distortions}
\end{figure}

As described in the introduction, LaMnO$_3$ exhibits a strongly
distorted perovskite structure with space group
$Pnma$.\cite{Elemans_et_al:1971} The deviation of this structure from
the perfect cubic perovskite structure (with space group $Pm\bar{3}m$)
can be decomposed into the following three contributions:
\begin{enumerate}
\item[(i)]{A staggered (checkerboard-like) $Q^x$-type JT distortion of
  the oxygen octahedra within the $x$-$y$ plane, with the long and
  short axes of neighboring octahedra alternating between the $x$ and
  $y$ directions (see Fig.~\ref{distortions}). This JT distortion
  leads to a doubling of the unit cell compared to the ideal cubic
  perovskite structure, with new ``in-plane'' lattice vectors $\vec{a}
  = a_0 (\hat{x} - \hat{y})$ and $\vec{b} = a_0 (\hat{x} + \hat{y})$,
  where $a_0$ is the lattice constant of the original (undistorted)
  perovskite structure. Identical $x$-$y$ planes are stacked on top of
  each other along the $z$ direction. The resulting symmetry is
  tetragonal.}
\item[(ii)]{``GdFeO$_3$-type'' rotations (tilting) of the oxygen
octahedra, leading to an additional doubling of the unit cell along
the $z$ direction, with the new lattice vector $\vec{c} =
2a_0\hat{z}$, and a reduction to orthorhombic $Pnma$ symmetry.}
\item[(iii)]{Displacements of the La cations, and a deformation
  (strain) of the parallelepiped formed by the lattice vectors
  $\vec{a}$, $\vec{b}$, and $\vec{c}$, consistent with the
  orthorhombic crystal class.}
\end{enumerate}
We expect that the internal distortions of the oxygen
network, i.e. components (i) and (ii) described above, have the
largest effect on the $e_g$ bands of LaMnO$_3$ via the ligand-field
splitting, whereas the influence of the lattice strain and of the La
displacements, i.e. component (iii), can be neglected. We test the
validity of this hypothesis in Sec.~\ref{sec:struc_test}.

In the following, except for the test calculations presented in
Sec.~\ref{sec:struc_test}, we therefore consider only components (i)
and (ii). This means that we use the ``pseudo-cubic'' lattice vectors
$\vec{a}$, $\vec{b}$, and $\vec{c}$ defined above, discarding any
orthorhombic strain, and we place the La cations on their ideal
positions, corresponding to the perfect cubic perovskite structure.
For the internal distortion of the oxygen network, we use the
experimental data obtained in Ref.~\onlinecite{Norby_et_al:1995},
which we decompose into the pure JT distortion and the GdFeO$_3$-type
distortion, as described below. Furthermore, we use a ``cubic''
lattice constant $a_0$ = 3.9345~\AA, which results in the same volume
per formula unit as in the experimentally observed
structure.\cite{Norby_et_al:1995}

\begin{table}
\caption{Decomposition of the experimentally observed distortion of
  the oxygen network into JT component and GdFeO$_3$-type (GFO)
  rotations. The upper four lines contain the Wyckoff positions of the
  two inequivalent oxygen sites O1 $(4c)$ and O2 $(8d)$ in the
  experimental structure (Ref.~\onlinecite{Norby_et_al:1995}), in the
  ideal cubic perovskite structure (note that these values correspond
  to a quadrupled unit cell), and our decomposition in pure JT and
  pure GdFeO$_3$-type components. $\Delta \vec{r}$, $\Delta
  \vec{r}_\text{JT}$, and $\Delta \vec{r}_\text{GFO}$ represent the
  corresponding full experimental distortion, and its decomposition
  into pure JT and GFO-type distortion, respectively. $x$, $y$, and
  $z$ are the coordinates with respect to the orthorhombic lattice
  vectors.}
\label{tab:pos}
\begin{ruledtabular}
\begin{tabular}{l|ccc|ccc}
 & & O1 $(4c)$ & & & O2 $(8d)$ & \\
 & $x$ & $y$ & $z$ & $x$ & $y$ & $z$ \\
\hline
Exp. (Ref.~\onlinecite{Norby_et_al:1995}) & -0.0733 & -0.0107 & 0.25 & 0.2257 & 0.3014 & 0.0385 \\
Ideal & 0.0 & 0.0 & 0.25 & 0.25 & 0.25 & 0.0 \\
JT & 0.0 & 0.0 & 0.25 & 0.2636 & 0.2636 & 0.0 \\
GFO & -0.0733 & -0.0107 & 0.25 & 0.2122 & 0.2879 & 0.0385 \\
\hline
$\Delta \vec{r}$ & -0.0733 & -0.0107 & 0.0 & -0.0243 & 0.0514 & 0.0385 \\
$\Delta \vec{r}_\text{JT}$ & 0.0 & 0.0 & 0.0 & 0.0136 & 0.0136 & 0.0 \\
$\Delta \vec{r}_\text{GFO}$ & -0.0733 & -0.0107 & 0.0 & -0.0379 & 0.0379 & 0.0385 
\end{tabular}
\end{ruledtabular}
\end{table}

Table~\ref{tab:pos} lists the Wyckoff positions for the two
inequivalent oxygen sites O1 $(4c)$ and O2 $(8d)$ in the
experimentally determined $Pnma$ structure,\cite{Norby_et_al:1995} and
how we decompose the corresponding structural distortion in the pure
JT component (i) and the GdFeO$_3$-type distortion (ii). The
decomposition is such that $\Delta \vec{r} = \Delta \vec{r}_\text{JT}
+ \Delta \vec{r}_\text{GFO}$ and $\Delta \vec{r}_\text{JT}$ is
orthogonal to $\Delta \vec{r}_\text{GFO}$, where $\Delta \vec{r}$,
$\Delta \vec{r}_\text{JT}$, and $\Delta \vec{r}_\text{GFO}$ are the
full experimental distortion and its decomposition into pure JT and
GdFeO$_3$-type distortion, respectively. Since the Wyckoff coordinates
$x$, $y$, and $z$ can be directly interpreted as the coordinates
relative to the lattice vectors $\vec{a}$, $\vec{b}$, and $\vec{c}$,
it follows that in the purely JT distorted structure each oxygen anion
O2 is displaced by $|\Delta \vec{r}_\text{JT}(\text{O2})| = | \Delta
x_\text{JT}(\text{O2})\,\vec{a} + \Delta
y_\text{JT}(\text{O2})\,\vec{b} + \Delta
z_\text{JT}(\text{O2})\,\vec{c}\,| = 0.1070$\,\AA. According to
Eq.~(\ref{JT_amp}) this corresponds to a JT amplitude of $Q^x_0 =
0.1513 \text{\AA} = 0.0385\,a_0$.

\section{Discussion of results}
\label{sec:results}

\subsection{Test of structural decomposition}
\label{sec:struc_test}

In the previous section we stated that only components (i) and (ii),
i.e. the internal distortion of the oxygen network, are important for
the $e_g$ bands in LaMnO$_3$, and that the lattice strain as well as
the displacements of the La cations are negligible. In order to test
this hypothesis, we now compare the LSDA band-structure calculated for
the full experimental structure of Ref.~\onlinecite{Norby_et_al:1995}
with the one calculated for the slightly simplified structure
described above, where the lattice strain and the La displacements are
set to zero, while the internal coordinates of the oxygen anions are
the same as observed experimentally.

\begin{figure}
\includegraphics*[width=\columnwidth]{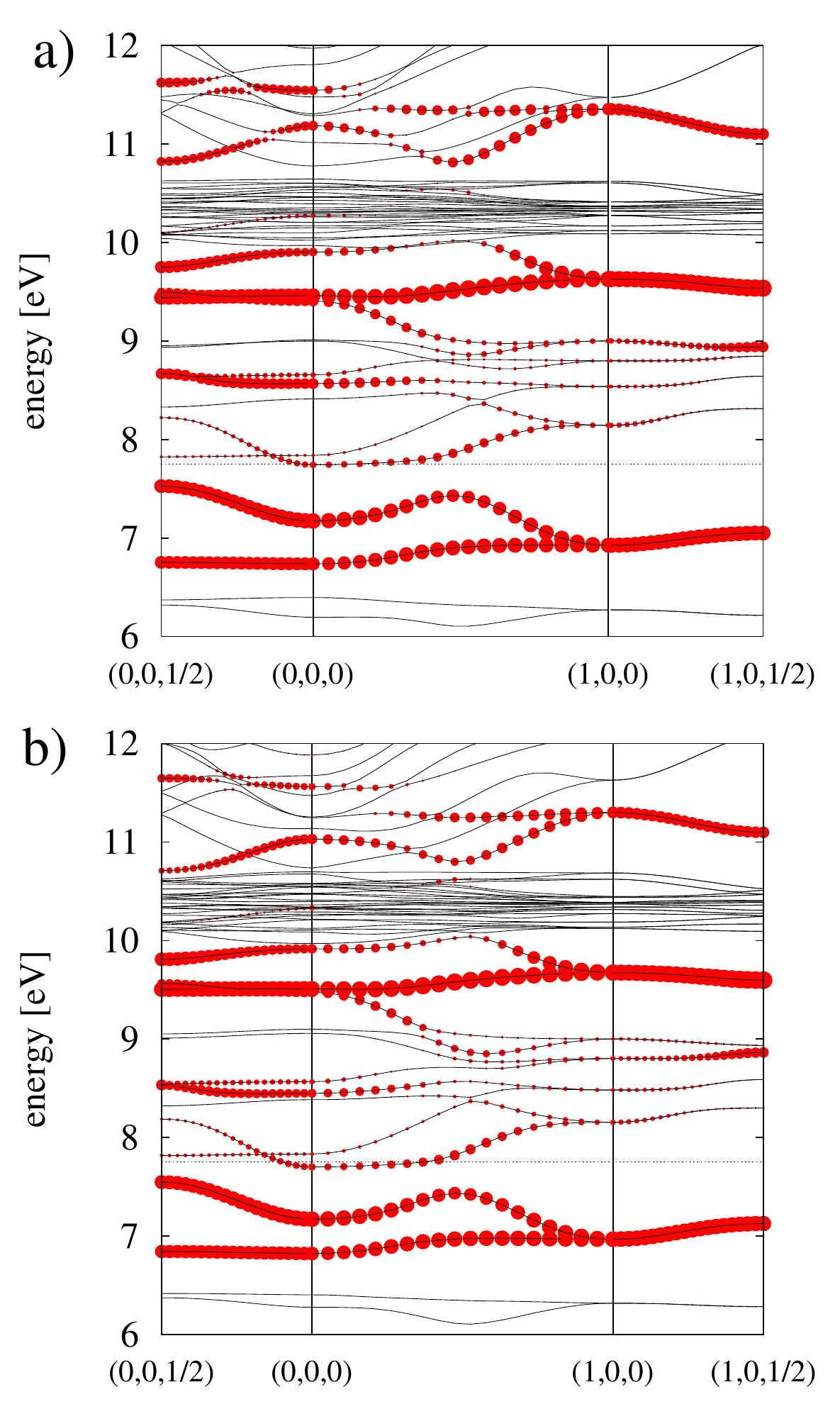}
\caption{LSDA energy bands (thin lines) along high symmetry directions
  of the orthorhombic Brillouin zone calculated for a) the exact
  experimental structure, and b) for the simplified structure with no
  orthorhombic strain and the La cations on their ideal
  positions. Both calculations are done for A-type antiferromagnetic
  ordering. The thick dots indicate the amount of Mn $e_g$ character
  in the corresponding Bloch functions. The dashed horizontal line at
  $\sim$ 7.75\,eV indicates the Fermi energy. In b) the high symmetry
  $k$-points are given in cartesian coordinates and in units of
  $\pi/a_0$, in a) the corresponding $k$-points are labeled
  identically but correspond to the slightly strained reciprocal
  lattice of the experimental structure. The two cases are nearly
  indistinguishable.}
\label{fig:strain-effect}
\end{figure}

The corresponding LSDA band-structures in the energy range of the Mn
$e_g$ bands calculated for A-type antiferromagnetic ordering are shown
in Fig.~\ref{fig:strain-effect} along certain high symmetry directions
of the orthorhombic Brillouin zone. The Mn $e_g$ bands are visualized
by dots along the bands, with the radius of the dots proportional to
the amount of Mn $e_g$ character contained in the corresponding
Bloch-function. It is clearly seen that the band-structures obtained
for the fully experimental structure and for the simplified structure
with only the oxygen distortions included are nearly
indistinguishable, with only small deformations of the energy bands
resulting from the orthorhombic strain and the La displacements. This
validates our initial hypothesis, and in the following we therefore
analyze only the effect of the internal structural distortion of the
oxygen octahedra on the dispersion of the $e_g$ bands.

We point out that by setting the lattice strain to zero we also
neglect any homogeneous $Q^z$-type JT distortion. The good agreement
between the two band-structures shown in Fig.~\ref{fig:strain-effect}
thus also indicates that there is no noticeable effect of $Q^z$ on the
electronic band-structure of LaMnO$_3$ in its experimental crystal
structure.

The simplified structure that gives rise to the LSDA band-structure
shown in the bottom part of Fig.~\ref{fig:strain-effect} results from
the superposition of distortions (i) and (ii) (described above) of the
oxygen network. In the following we will first establish the $e_g$
band-structure of LaMnO$_3$ in the ideal cubic perovskite structure
and then separately introduce either the JT distortions, component
(i), or the GdFeO$_3$-type rotations, component (ii), and analyze the
corresponding effects on the $e_g$ bands.

\subsection{Cubic structure}
\label{sec:cubic}

\begin{figure}
\includegraphics[width=\columnwidth]{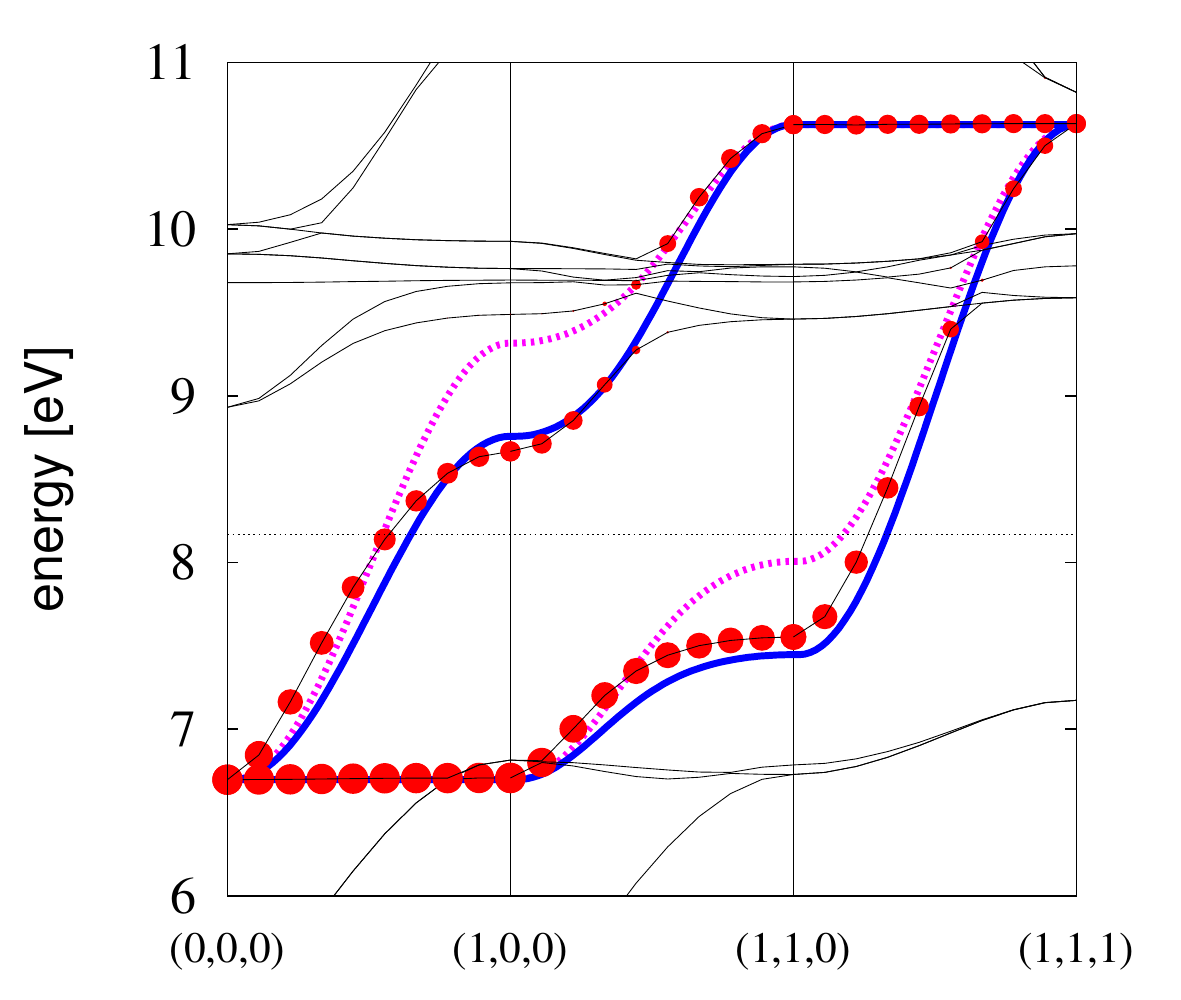}
\caption{Majority spin bands for cubic ferromagnetic LaMnO$_3$. The
LSDA band-structure is represented by thin lines, with the dots
indicating the amount of Mn $e_g$ character in the corresponding Bloch
function. The thick broken line is the TB fit with only nearest
neighbor hopping, whereas the thick solid line represents the TB fit
including both nearest and next nearest neighbor hopping. The thin
dashed horizontal line at $\sim$ 8.2\,eV indicates the Fermi
energy. The high symmetry k-points are given in cartesian coordinates
and in units of $\pi/a_0$.}
\label{cubFM}
\end{figure}

Fig.~\ref{cubFM} shows the calculated LSDA energy dispersion around
the Fermi-level, calculated for ferromagnetic cubic LaMnO$_3$ in the
undistorted cubic perovskite structure with $a_0$ = 3.9345~\AA. Only
the majority spin bands are shown.  Within the TB model defined in
Eq.~(\ref{model}), the Hund's coupling simply splits the spin majority
and minority bands rigidly by $\Delta E = 2J$ for a ferromagnetic
arrangement of the $t_{2g}$ core spins, and we therefore discuss only
the majority spin bands in the following. The Mn $e_g$ bands are again
visualized by the dots along the bands in Fig.~\ref{cubFM}. It is
evident that even though the Mn $e_g$ bands are intersected by other
bands, the $e_g$ dispersion can be nicely traced along the dots. The
$e_g$ bands are about half-filled, as expected from the formal
electron configuration $t_{2g}^3\,e_g^1$ of the Mn$^{3+}$ ion. The
bands at $\sim$ 7\,eV and lower, just touching the lower range of the
$e_g$ bands, are the filled majority Mn $t_{2g}$ bands. The weakly
dispersive bands slightly below 10\,eV that intersect the Mn $e_g$
bands correspond to the La 4$f$ states, and the strongly dispersive
unoccupied bands above the Mn $e_g$ manifold have predominantly La $d$
character.

The thick lines in Fig.~\ref{cubFM} correspond to fits of the nearest
and next nearest neighbor TB models for the $e_g$ bands. The nearest
neighbor hopping parameter $t = 0.655$\,eV is determined from the full
$e_g$ bandwidth $W = 3.928$\,eV$ = 6t$. The next nearest neighbor
hopping parameter $t'$ is obtained in the following way: In the next
nearest neighbor model the width of the energy dispersion of the upper
$e_g$ band between $k$-points $\Gamma=(0,0,0)$ and $X=(1,0,0)$ is
equal to $\Delta E^{(2)}_{\Gamma X} = 4t-16t'$, whereas the dispersion
width of the lower band between $X$ and $M=(1,1,0)$ is equal to
$\Delta E^{(1)}_{XM} =
2t-16t'$.\cite{footnote}\nocite{Bradley/Cracknell:Book} The
corresponding energy differences obtained from the LSDA calculation
are $\Delta E^{(2)}_{\Gamma X} = 0.851$\,eV and $\Delta E^{(1)}_{XM} =
1.965$\,eV, leading to $t' = 0.041$\,eV or $t' = 0.029$\,eV,
respectively (and using the previously obtained $t$=0.655\,eV). An
average value of $t' = 0.035$\,eV is used for the TB fit in
Fig.~\ref{cubFM} and in the remaining part of this paper.

It becomes clear from Fig.~\ref{cubFM} that the simple nearest neighbor
TB model cannot reproduce the LSDA dispersion very well, whereas the
next nearest neighbor TB model leads to a very good description of the
energy dispersion for all $k$-points.

We point out that a nonmagnetic LDA calculation results in a low-spin
electron configuration and the loss of the $t_{2g}$ core spin. Indeed,
if we perform a nonmagnetic LDA calculation, the $e_g$ bands are empty
and higher in energy compared to the oxygen $p$ levels, which in
accordance with Eq.~(\ref{t_eff}) results in a reduced bandwidth of
3.546\,eV, corresponding to a nearest neighbor hopping amplitude of
$t=0.591$\,eV. A nonmagnetic LDA calculation is thus not necessarily a
good representation of the electronic structure of the paramagnetic
phase, and we therefore use the ferromagnetic state as the starting
point for the model analysis. In general, this shows that an LDA+DMFT
treatment of LaMnO$_3$ based on a nonmagnetic LDA calculation, such as
the one presented in Ref.~\onlinecite{Yamasaki_et_al:2006}, leads to a
slight underestimation of the electron hopping.

\begin{figure}
\includegraphics[width=\columnwidth]{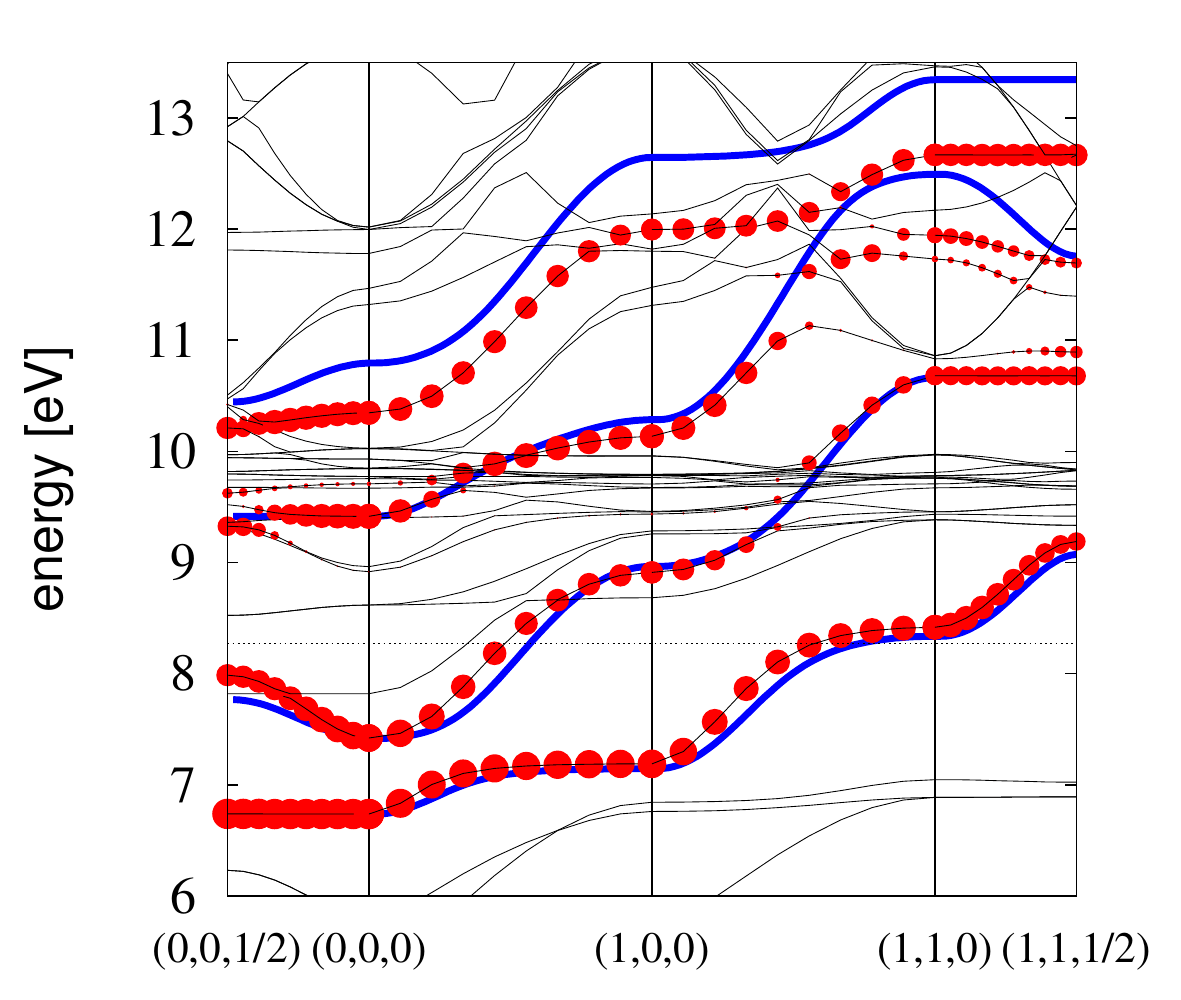}
\caption{Band-structure for cubic LaMnO$_3$ with A-type
  antiferromagnetic order. The LSDA band-structure is represented by
  thin lines, with the dots indicating the amount of Mn $e_g$
  character in the corresponding Bloch function. The thick line
  represents the TB fit including both nearest and next nearest
  neighbor hopping. The thin dashed horizontal line at $\sim$ 8.3\,eV
  indicates the Fermi energy. The high symmetry k-points are given in
  cartesian coordinates and in units of $\pi/a_0$.}
\label{cubAFM}
\end{figure}

Next we investigate the influence of A-type antiferromagnetic
order. Fig.~\ref{cubAFM} shows the calculated LSDA band-structure for
this case. Note that the underlying crystal structure is still perfect
cubic perovskite. Again, the $e_g$ character of the bands is
visualized by the dots. The thick lines corresponds to the fit within
our antiferromagnetic next nearest neighbor TB model with the hopping
parameters obtained from the ferromagnetic case. Due to the doubling
of the magnetic unit cell the number of bands is also doubled. A
Hund's-rule parameter $J = 1.340$\,eV is obtained from the energy
splitting at the $\Gamma$ point between the two bands at 6.7\,eV and
9.4\,eV, which show no dispersion along $\overline{\Gamma A}$
($A=(0,0,1/2)$). This splitting is exactly equal to $2 J$ in the TB
model. The value $J=1.340$\,eV is within 2.4\,\% of the value
$J=1.309$\,eV obtained from the ferromagnetic band-structure as the
splitting between majority and minority spin states at the $\Gamma$
point (not shown).

Fig.~\ref{cubAFM} shows that the two lowest $e_g$ bands are described
very well by the antiferromagnetic TB model. The upper two bands show
some deviations from the model, especially in the high energy
region. This is an inevitable result of the description within a pure
$d$ band model. As described in Sec.~\ref{sec:intro}, the ``true''
hopping is mediated by the oxygen $p$ orbitals and therefore the $e_g$
dispersion depends on the energetic distance from the oxygen $p$
levels (see Eq.~(\ref{t_eff})). This leads to a slight overestimation
of the energy dispersion for the high energy states in the pure $d$
model.  The same effect can also be observed in the ferromagnetic
case: Due to their higher energy relative to the oxygen $p$ states,
the bandwidth of the $e_g$ minority spin bands is smaller than for the
corresponding majority spin bands. The nearest neighbor hopping
parameter corresponding to the minority spin bands in the
ferromagnetic case is $t=0.548$\,eV. In the following we use the value
$t=0.655$\,eV, corresponding to the majority spin bands in the
ferromagnetic configuration, since this value is representative for
the $e_g$ bands close to the Fermi level which determine the important
low energy behavior in manganite systems.

\begin{figure}
\includegraphics[width=0.9\columnwidth]{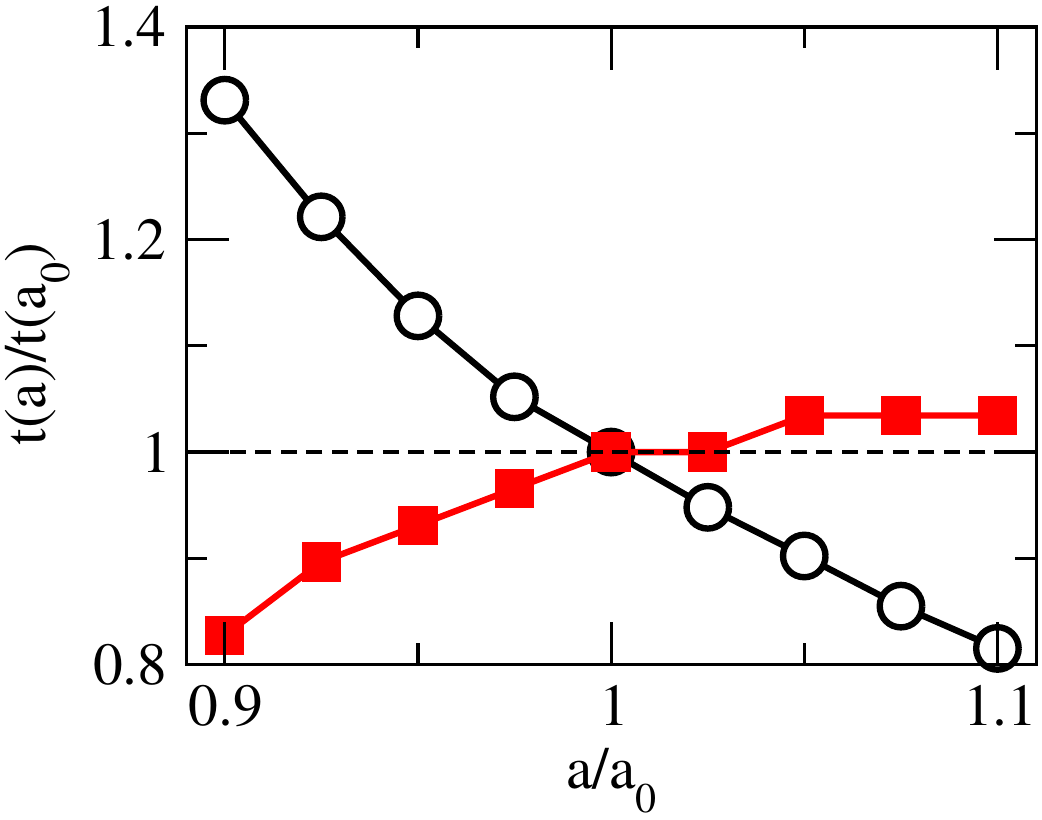}
\caption{Variation of hopping amplitudes with lattice constant. Open
  circles correspond to the nearest neighbor hopping $t$ and filled
  squares correspond to the next nearest neighbor hopping $t'$. $a_0$
  indicates the lattice constant derived from the experimental unit
  cell volume.}
\label{volume}
\end{figure}

Fig.~\ref{volume} shows the dependence of both nearest and next
nearest hopping parameters on the lattice constant $a$. The nearest
neighbor hopping $t$ decreases with increasing Mn-O bond length,
whereas the next nearest neighbor hopping $t'$ shows a slight
increase. This somewhat unexpected behavior of $t'$ results from the
fact that the energy difference between oxygen $p$ and Mn $d$ states
decreases with increasing volume, and therefore counteracts the effect
of the reduced overlap integrals for larger lattice constants.

\subsection{Purely JT distorted structure}

We now address the effect of the JT distortion of the oxygen octahedra
on the $e_g$ bands in LaMnO$_3$. As described in Sec.~\ref{sec:struc},
we consider only the internal distortions of the oxygen network. This
means that we displace the oxygen anions relative to the cubic
structure according to the decomposition of the Wyckoff positions
described in Table~\ref{tab:pos} (line ``JT''). This results in the
staggered arrangement of JT distorted oxygen octahedra shown in
Fig.~\ref{distortions}b, with the same JT amplitude $Q^x_0$ as in the
experimental structure.

\begin{figure}
\includegraphics[width=\columnwidth]{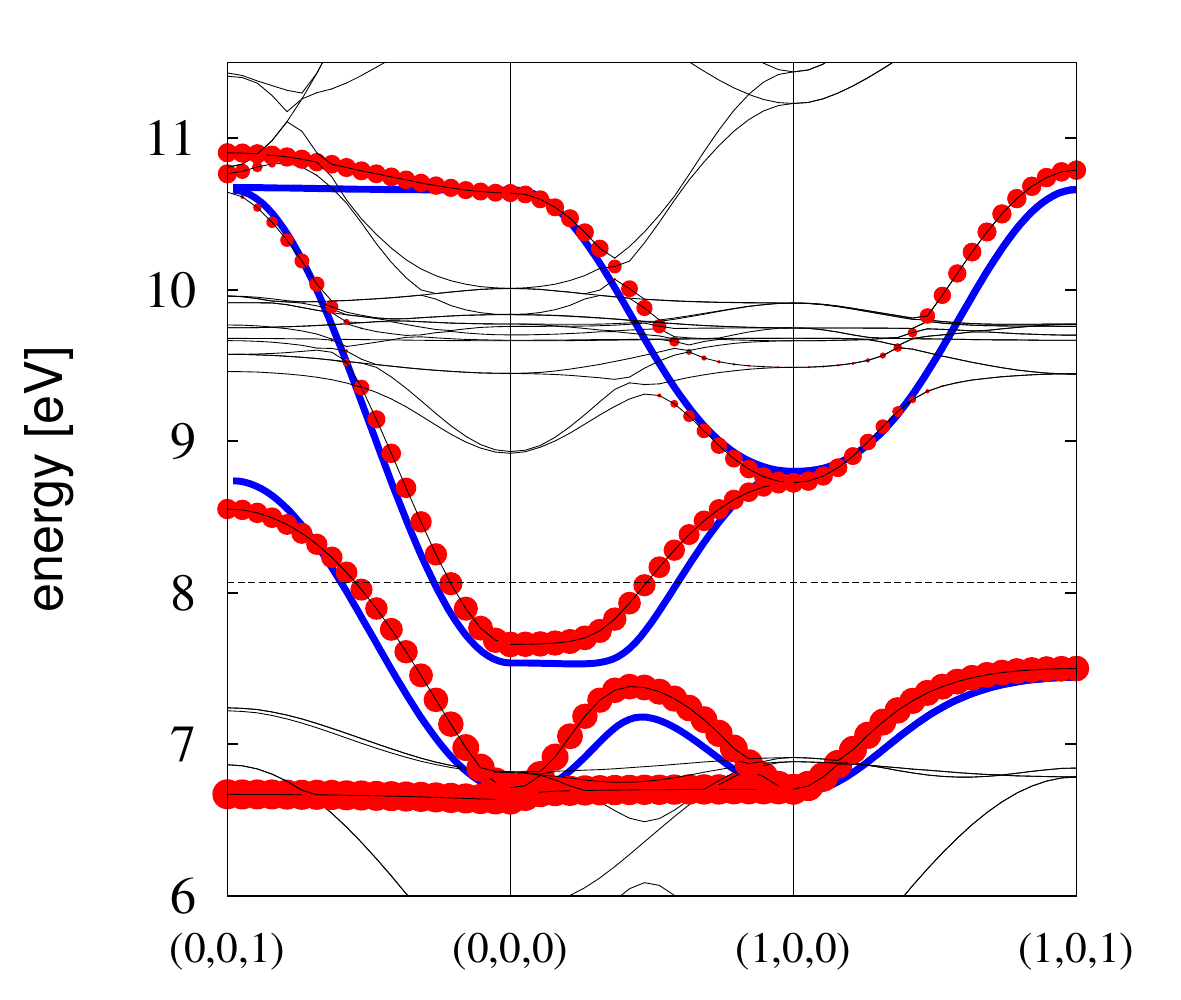}
\caption{Majority-spin band-structure for ferromagnetic LaMnO$_3$ in
  the purely JT distorted structure (see Fig.~\ref{distortions}b),
  where the unit cell is doubled in the $x$-$y$-plane. The high
  symmetry k-points are given with respect to the cartesian coordinate
  system defined by the reciprocal lattice vectors of the undistorted
  cubic structure. The units are $\pi/a_0$. The LSDA band-structure is
  represented by thin lines with the dots indicating the amount of Mn
  $e_g$ character in the corresponding Bloch function. The thick line
  represents the TB fit and the thin dashed horizontal line at $\sim$
  8.1\,eV indicates the Fermi energy. }
\label{onlyJT}
\end{figure}

Fig.~\ref{onlyJT} shows the calculated LSDA majority spin bands for
LaMnO$_3$ in the purely JT distorted structure with ferromagnetic spin
ordering. Due to the unit cell doubling within the $x$-$y$ plane
compared to the cubic structure, the two bands corresponding to the
$k$-points between $X=(1,0,0)$ and $M=(1,1,0)$ in the ferromagnetic
cubic case (see Fig.~\ref{cubFM}) are ``back-folded'' between the
$k$-points $U=(1,0,0)$ and $\Gamma=(0,0,0)$ in the tetragonal
Brillouin zone. In addition, the two bands between $M=(1,1,0)$ and
$R=(1,1,1)$ in Fig.~\ref{cubFM} now correspond to the two upper bands
between $\Gamma=(0,0,0)$ and $Z=(0,0,1)$ in Fig.~\ref{onlyJT}, and the
two bands between $U=(1,0,0)$ and $R=(1,0,1)$ in Fig.~\ref{onlyJT} are
now twofold degenerate. It can be seen that the level splitting
between the two original $e_g$ states at the $\Gamma$ point (at
$\sim$6.7\,eV in Fig.~\ref{onlyJT}), which are degenerate in the cubic
case, is very small ($\sim$ 0.08\,eV), and that the main effect of the
JT distortion is to remove the band crossing between $\Gamma$ and
$U=(1,0,0)$, resulting from the simple ``back-folding'' of the cubic
band-structure due to the unit cell doubling.

To obtain the value of the JT coupling constant $\lambda$ within our
TB model, we first determine the $k$-point of the band crossing
between $\Gamma$ and $U$ for the case of zero JT distortion. We then
determine $\lambda Q^x_0$ in the model by fitting the splitting at
this $k$-point to the corresponding splitting obtained from the LSDA
calculation. In this way we obtain a value of $\lambda Q^x_0 =
0.248$\,eV, corresponding to $\lambda = 1.639\,$eV/\AA, since $Q^x_0 =
0.1513$\,\AA\ (see Sec.~\ref{sec:struc}). Alternatively we can also
fit the small splitting of the two lowest $e_g$ bands at the $\Gamma$
point by numerically adjusting the JT coupling. In this way we find
$\lambda Q^x_0 = 0.289$\,eV, corresponding to $\lambda =
1.910\,$eV/\AA, which is within 17\,\% of the value obtained
above. This shows that the extracted coupling strength does not depend
critically on the fitting procedure.

The energy dispersion calculated within the TB model using the hopping
amplitudes $t=0.655$\,eV and $t'=0.035$\,eV obtained for the cubic
structure and the JT coupling constant $\lambda=1.639$\,eV/\AA\
obtained in the way described above is compared to the full LSDA
band-structure in Fig.~\ref{onlyJT}. It can be seen that the
dispersion of the $e_g$-projected bands is well described within the
TB model. Some deviations occur close to the lifted band crossing
between $\Gamma$ and $U = (1,0,0)$. These deviations are most likely
caused by the asymmetry of the Mn-O bonds, which is neglected in the
effective $e_g$ TB model. The good overall quality of the TB fit shows
that the TB hopping amplitudes are not affected by the presence of the
JT distortion. This indicates that the model description with the
assumed local coupling of the JT distortion to the $e_g$ levels is
justified.

\begin{figure}
\includegraphics[width=0.9\columnwidth]{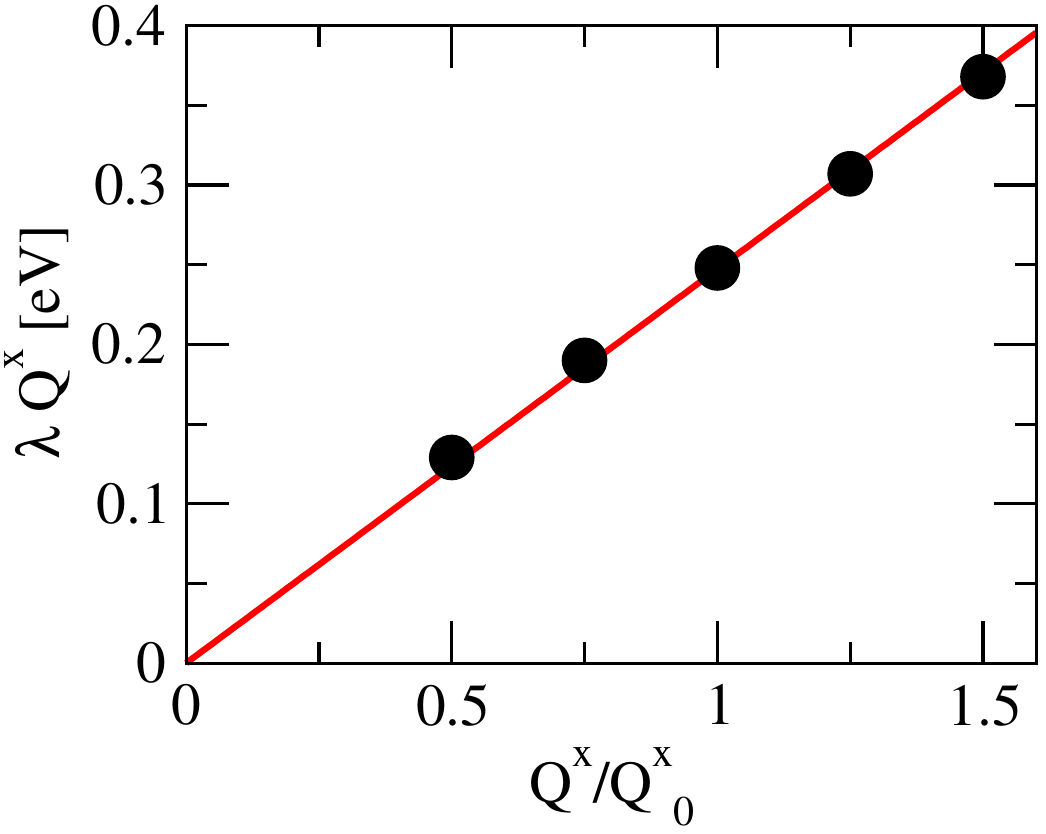}
\caption{Magnitude of $\lambda Q^x$ obtained by fitting the TB model
  to LSDA calculations with different amplitudes of the JT
  distortion. $Q^x_0$ indicates the magnitude of the JT distortion
  found in the experimental structure. Filled circles are the results
  of the actual calculations, whereas the straight line corresponds to
  $\lambda Q^x = 0.247\,\text{eV}\ Q^x/Q^x_0$.}
\label{JT_linear}
\end{figure}

To test whether the linear form of the coupling term within our model
is consistent with the density functional theory calculation, and to
further test our procedure for obtaining $\lambda Q^x$ from the LSDA
results, we perform additional calculations with different amplitudes
of the JT distortion. The results are presented in
Fig.~\ref{JT_linear}, where the JT energy $\lambda Q^x_0$ is
determined by fitting the band-splitting between $\Gamma$ and
$U=(1,0,0)$. We note that $\lambda Q^x$ on the ordinate in
Fig.~\ref{JT_linear} should be regarded as the model parameter that we
obtain by our fitting procedure, whereas the abscissa $Q^x/Q^x_0$
characterizes the input structure for our LSDA calculation (relative
to the experimentally observed JT distortion $Q^x_0$). It is evident
that the dependence of $\lambda Q^x$ on the input distortion is nearly
perfectly linear, which indicates the good quality of our fit and the
adequacy of the linear coupling term within the model.

It becomes clear from Fig.~\ref{onlyJT} that the JT distortion in the
experimentally observed structure of LaMnO$_3$ has only a weak effect
on the dispersion along $k_z$. This is complementary to the effect of
the A-type magnetic order, which strongly suppresses the electron
hopping along this direction. The insulating band-structure obtained
in LSDA for the fully distorted structure with A-type
antiferromagnetic order is therefore a combined effect of both the
staggered JT distortion within the $x$-$y$ plane and the A-type
antiferromagnetism. To achieve an insulating state within LSDA solely
due to the JT distortion (i.e. for the ferromagnetic case), would
require an unrealistically large JT amplitude. Within our TB model, a
value of $\lambda Q^x > 1.1$\,eV, i.e. more than four times the JT
distortion of the experimental structure, is required to open up an
energy gap. This is due to the the large value of the hopping $t$ and
the fact that for staggered JT order $H_\text{JT}$ does not commute
with $H_\text{kin}$. The fact that the JT distortion alone is not
enough to stabilize an insulating state in LaMnO$_3$ has also been
pointed out in
Refs.~\onlinecite{Ahn/Millis:2000,Yin/Volja/Ku:2006,Yamasaki_et_al_LMO:2006}.

\begin{figure}
\includegraphics[width=\columnwidth]{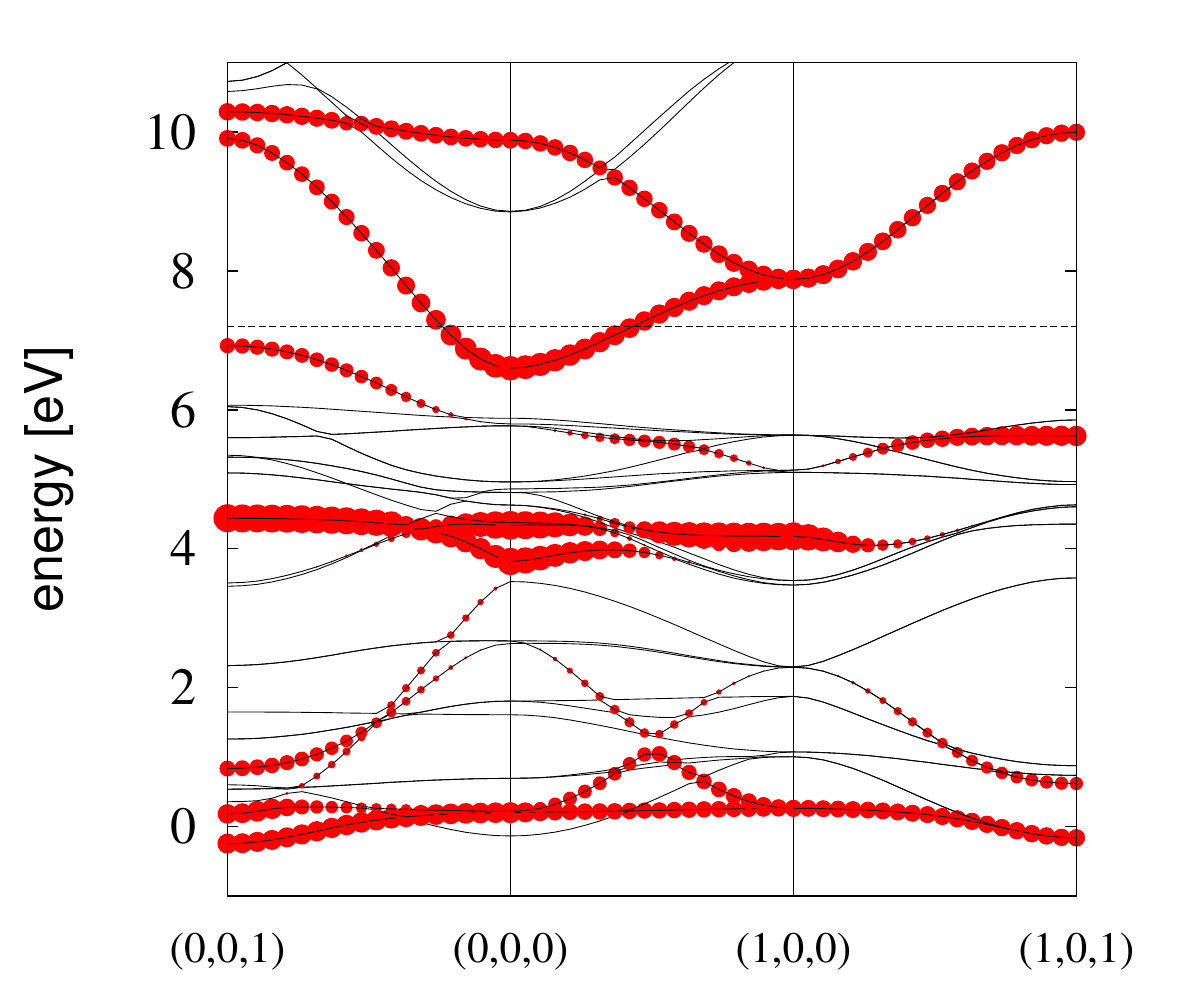}
\caption{Majority-spin band-structure for ferromagnetic LaMnO$_3$ in
  the purely JT distorted structure calculated using the LSDA+$U$
  method with $U_\text{eff} = 7.12$\,eV ($U=8$\,eV and ${\cal J}^H =
  0.88$\,eV). The high symmetry $k$-points are given in cartesian
  coordinates and in units of $\pi/a_0$. The Kohn-Sham band-structure
  is represented by thin lines with the dots indicating the amount of
  Mn $e_g$ character in the corresponding Bloch function. The thin
  dashed horizontal line at $\sim$7.2\,eV indicates the Fermi
  energy. The overlapping bands in the energy range between 0\,eV and
  6\,eV have mixed O\,2$p$-Mn\,3$d$ character.}
\label{ldau}
\end{figure}

One possibility for explaining the insulating character of LaMnO$_3$,
as noted by previous authors,\cite{Ahn/Millis:2000} is that
electron-electron interactions beyond LSDA increase the effective JT
splitting, thereby stabilizing the insulating state. To address this
we have performed additional LSDA+$U$ calculations (which will be
discussed in detail elsewhere). In the LSDA+$U$ method, the
interactions between the $d$ states of the transition metal cations
are treated explicitly on the level of a mean-field Hubbard
model.\cite{Anisimov/Aryatesiawan/Liechtenstein:1997} Thereby, a
parameter $U$ represents the strength of the (screened) on-site
Coulomb repulsion between the $d$ electrons and a parameter ${\cal
J}^H$ represents the Hund's coupling. In our LSDA+$U$ calculations we
use a slightly simplified approach where only $U_\text{eff}=U-{\cal
J}^H$ enters.\cite{Dudarev_et_al:1998}

We expect that the on-site Coulomb repulsion enhances the effect of
the JT distortion and therefore drives the system towards an
insulating state as the value of $U$ is increased. However, the
calculated LSDA+$U$ band-structure for the purely JT distorted case
with $Q^x = Q^x_0$ and ferromagnetic spin order stays metallic even
for a rather large Hubbard parameter of $U=8$\,eV (see
Fig.~\ref{ldau}). The reason for this is the following: The use of the
LSDA+$U$ method leads to a strong downward energy shift of the
occupied part of the $e_g$ bands and also to a moderate overall
downshift of the $d$ states as a whole. As a result of the latter, the
higher-lying, mostly unoccupied $e_g$ bands in LaMnO$_3$ move somewhat
closer to the O 2$p$ bands, which are located in the energy range
between 0\,eV and 6\,eV and are not shifted within the LSDA+$U$
method. Furthermore, the strong downward shift of the occupied $d$
bands results in an energetic overlap and therefore strong
hybridization between the occupied $e_g$ states and the O 2$p$
bands. In this case the simple two-band TB model is not applicable any
more, and some mixed $p$-$d$ bands, which extend above the original
top of the O 2$p$ states at 6\,eV, reach above the Fermi level. Since
the LSDA+$U$ method corrects only for the Coulomb interactions between
the Mn $d$ states, it is not obvious whether this shift of the $d$
states relative to the O 2$p$ states is a real physical effect or
rather an artifact of the LSDA+$U$ method (see Sec.~\ref{sec:summary}
for a further discussion of this point). An exact experimental
determination of the energy separation between the Mn $d$ and O$p$
states in LaMnO$_3$ would provide further insight on this.

\subsection{GdFeO$_3$-type rotations}

Finally, we address the effect of the GdFeO$_3$-type oxygen octahedra
rotations. These distortions change the Mn-O-Mn bond angles and are
therefore expected to alter the magnitude of the hopping
amplitudes. In addition, due to the resulting symmetry lowering, this
distortion will enable hopping between orbitals that was either
symmetry-forbidden or negligibly small in the undistorted state.

\begin{figure}
\includegraphics[width=\columnwidth]{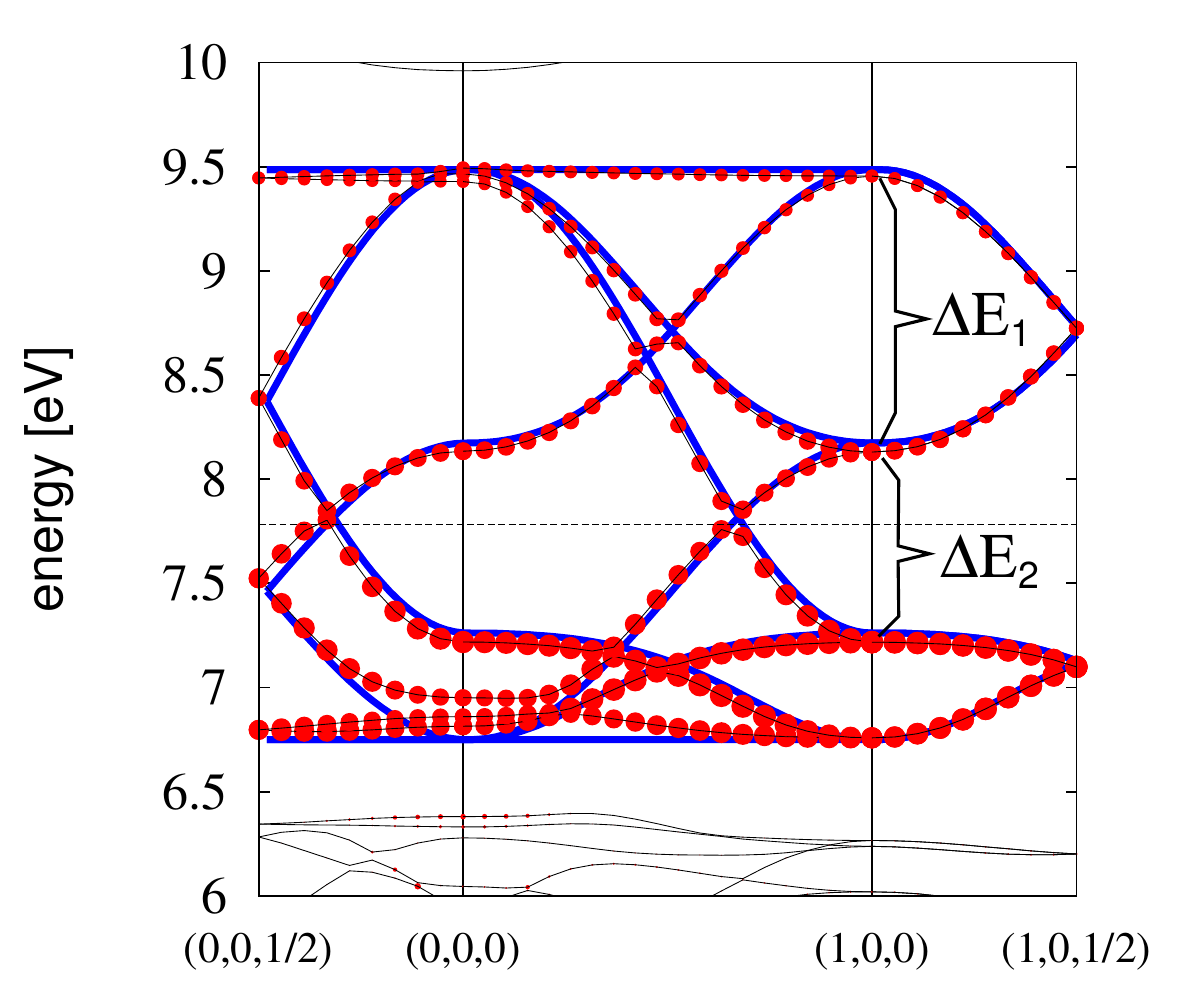}
\caption{LSDA energy bands (majority spin) for ferromagnetic LaMnO$_3$
  with the experimentally observed GdFeO$_3$-type tilting of the
  oxygen octahedra, but without JT distortion. The LSDA band-structure
  is represented by thin lines with the dots indicating the amount of
  Mn $e_g$ character in the corresponding Bloch function. The thick
  line represents the TB fit with reduced hopping amplitudes and the
  thin dashed horizontal line at $\sim$ 7.8\,eV indicates the Fermi
  energy. The high symmetry k-points are given in cartesian
  coordinates and in units of $\pi/a_0$. $\Delta E_1$ and $\Delta E_2$
  mark the energy differences plotted in Fig.~\ref{reduction}.}
\label{no_JT}
\end{figure}

Fig.~\ref{no_JT} shows the calculated LSDA energy bands for the
structure where only the GdFeO$_3$-type distortion, component (ii) in
Sec.~\ref{sec:struc}, is included, whereas the JT distortion,
component (i), is set to zero. The Wyckoff positions of the oxygen
anions for this configuration are listed in Table~\ref{tab:pos} (line
``GFO''). In this structure the Mn-O-Mn bond angles are reduced from
the ideal 180$^\circ$ to about 155$^\circ$. The thick line in
Fig.~\ref{no_JT} corresponds to a fit within the next-nearest neighbor
TB model with both nearest and next-nearest hopping amplitudes scaled
by a factor of 0.7 compared to the perfectly cubic case. The high
quality of the fit is striking, even though the LSDA band structure
shows some additional dispersion at the bottom of the $e_g$ bands
which is not accounted for in the TB model. It appears that, to a good
accuracy, the oxygen tilts can be incorporated in the model simply by
reducing the hopping amplitudes in an appropriate way without having
to include additional hopping parameters due to the lower symmetry.

\begin{figure}
\includegraphics[width=0.9\columnwidth]{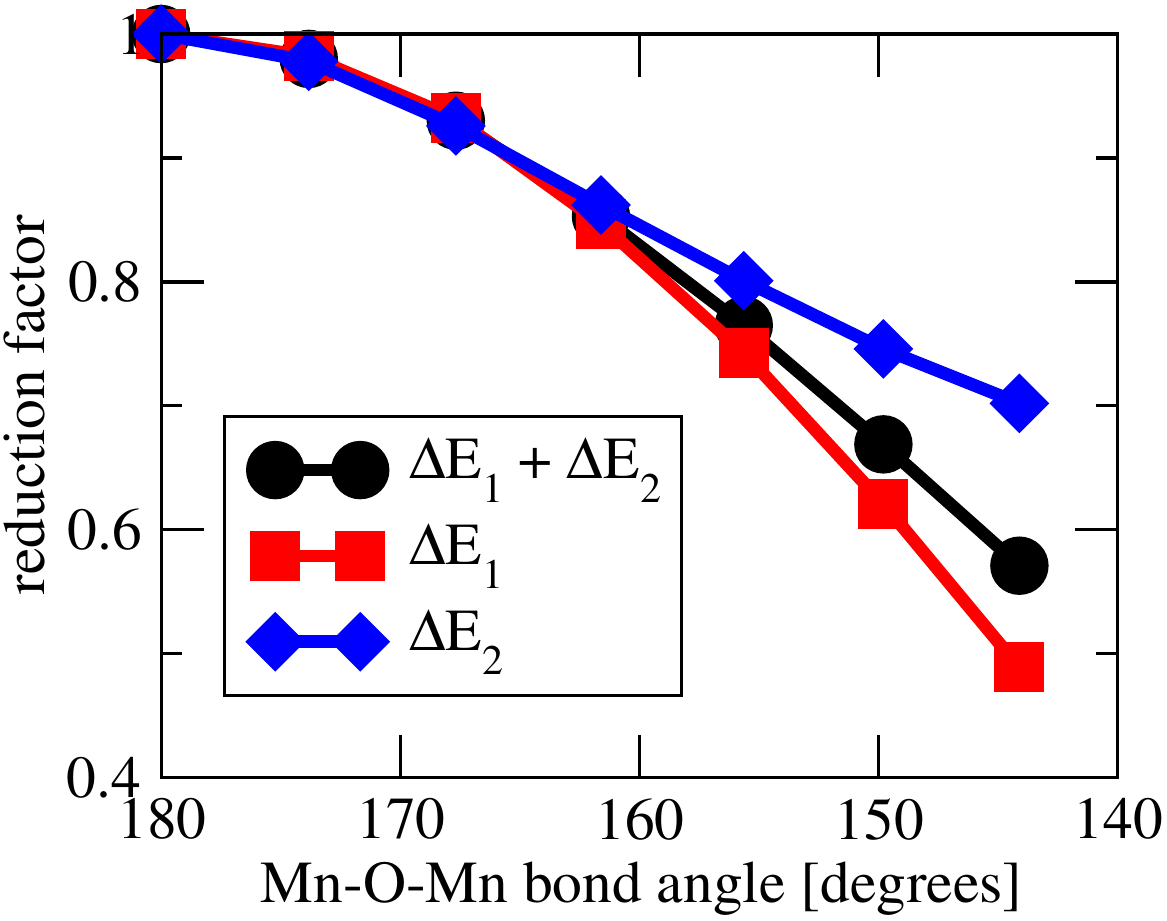}
\caption{Reduction of some characteristic energy differences at the
  $k$-point $U=(1,0,0)$ (see Fig.~\ref{no_JT}) for different
  amplitudes of the GdFeO$_3$-type distortion, which are indicated by
  the corresponding Mn-O-Mn bond angle.}
\label{reduction}
\end{figure}

To further quantify the reduction of the hopping amplitudes as a
result of the GdFeO$_3$-type distortion, we perform calculations for
different degrees of distortion, by scaling $\Delta
\vec{r}_\text{GFO}$ (see Table~\ref{tab:pos})
accordingly. Fig.~\ref{reduction} shows the resulting reduction of
some characteristic energy differences at the $k$-point $U=(1,0,0)$ as
a function of the Mn-O-Mn bond angle. If the GdFeO$_3$-type rotations
would lead to a simple scaling of the undistorted band-structure, all
the lines in Fig.~\ref{reduction} would fall on top of each other. It
can be seen that this is in fact a good approximation for Mn-O-Mn bond
angles down to $\sim$155$^\circ$, which corresponds to the
experimentally observed structure of LaMnO$_3$. For bond angles
smaller than 155$^\circ$ the band-structure starts deviating more
significantly from the cubic case due to new hopping paths that become
allowed in the distorted structure.

\section{Summary and conclusions}
\label{sec:summary}

In summary, we have shown, by comparing LSDA band-structure
calculations to simple TB models, that the relevant electronic states
in LaMnO$_3$ are well described by a model of $e_g$ orbitals with
nearest and next-nearest neighbor hoppings. We have quantified the
effect of changes in bond length (Fig.~\ref{volume}) and of the
octahedral rotations (Fig.~\ref{reduction}) on the hopping parameters,
and we find that for physically relevant values (bond angles $\gtrsim
155^\circ$) the GdFeO$_3$-type rotations significantly change the
value of the hopping parameters but do not invalidate the TB
description. Of particular importance is our finding that both the JT
lattice distortions and the Hund's rule coupling are quantitatively
accounted for by adding on-site interactions to the TB model,
\emph{without} changing the hopping parameters. In summary, these
results justify a TB plus interaction description of manganite
systems, and suggests more generally that such a description is useful
for transition metal oxides.

The parameters for nearest and next nearest neighbor hopping (defined
in Eqs.~(\ref{ekin}) and (\ref{hopping_start})-(\ref{hopping_end}))
which we obtain for the ideal cubic structure with lattice constant
$a_0 = 3.935$\,\AA\ are $t=0.655$\,eV and $t'=0.035$\,eV. The Hund's
rule and JT coupling constants (defined in
Eqs.~(\ref{hund})-(\ref{jt})) which we obtain from our analysis are
$J=1.34$\,eV and $\lambda = 1.64$\,eV/\AA. Our value of $J=1.34$\,eV
for the Hund's coupling is in excellent agreement with the exchange
splitting $2J=2.7$\,eV derived from x-ray absorption
measurements,\cite{Dessau/Shen:2000} and agrees well with previous
LSDA calculations.\cite{Ahn/Millis:2000,Yamasaki_et_al:2006} On the
other hand, the strength of the JT coupling $\lambda$ obtained in this
work is smaller than what has been concluded from x-ray absorption
measurements in Ref.~\onlinecite{Dessau/Shen:2000} It is also smaller
compared to various other values obtained previously from (mostly less
elaborate) fits of similar TB models to LSDA or LSDA+$U$
band-structure
calculations.\cite{Ahn/Millis:2000,Popovic/Satpathy:2000,Yin/Volja/Ku:2006,footnote:notation}

Popovic and Satpathy used a fitting procedure very similar to the one
presented here, and obtained a JT coupling strength of $\lambda =
2.8\,$eV/\AA\ (in our notation), compared to $\lambda = 1.64\,$eV/\AA\
obtained in the present work.\cite{Popovic/Satpathy:2000} We ascribe
this difference to the use of the atomic sphere approximation (ASA) in
the LSDA calculation of Ref.~\onlinecite{Popovic/Satpathy:2000}. In
the ASA the crystal volume is represented by an arrangement of
overlapping ``atomic spheres''.\cite{Andersen:1973} This overlap
introduces an error, which furthermore depends on the amplitude of the
JT distortion (since the JT distortion changes the overlap between the
atomic spheres) and thus can have a pronounced effect on $\lambda$.

Ahn and Millis used a TB model very similar to the one used in this
work, except that they didn't include the effect of next-nearest
neighbor hopping.\cite{Ahn/Millis:2000} They obtained a value of
$\lambda = 3.38\,$eV/\AA\ by simultaneously fitting 15 energies at 4
different high symmetry $k$-points to a previous LSDA calculation for
the fully distorted antiferromagnetically ordered case. It is not
obvious how sensitive such a simultaneous root mean square fit of all
the model parameters is to the exact value of $\lambda$, but we expect
that the neglect of next nearest neighbor hopping will lead to a
renormalization of the other parameters of the TB model in order to
account for the missing dispersion due to the next-nearest neighbor
hopping.

A value of $\lambda = 2.85\,$eV/\AA\ was obtained by Yin et
al.\cite{Yin/Volja/Ku:2006} by calculating the dependence of several
quantities on the amplitude of the JT distortion. This dependence was
first obtained from LSDA+$U$ calculations within a Wannier function
representation, and then compared to the corresponding results
calculated for a model Hamiltonian including electron-electron
interactions within the Hartree-Fock approximation. As in the case of
Ref.~\onlinecite{Ahn/Millis:2000} discussed above, it is not clear how
sensitive this simultaneous fit of all parameters in the model
Hamiltonian is to moderate changes in $\lambda$.


In contrast, the fitting procedure described in this work isolates the
effect of each term in the Hamiltonian (Eq.~(\ref{model})) and thus
allows to obtain each parameter independently from all others. In
particular, it becomes clear from our calculated band-structure shown
in Fig.~\ref{onlyJT}, that the JT distortion does not lead to a rigid
splitting of the $e_g$ bands, but that instead it has only subtle,
albeit rather important effects on band crossings at certain
$k$-points in the Brillouin zone.  The reason for this is that the JT
Hamiltonian $H_\text{JT}$ in Eq.~(\ref{jt}) for $Q^x\neq 0$ and
staggered order does not commute with the kinetic energy term in
Eq.~(\ref{ekin}), and that due to the relatively small value $\lambda
Q^x_0 \approx 0.25$\,eV the hopping energies are dominant, so that in
general the effect of the JT distortion is only visible as second
order shifts in the energy. In other words, at a generic $k$-point the
states picked out by the hopping term are not the eigenstates of the
JT distortion. This suggests that the straightforward interpretation
of peak splittings in the x-ray absorption spectra of
Ref.~\onlinecite{Dessau/Shen:2000} as a direct consequence of the JT
distortion is not necessarily justified.

Finally, our analysis enables us to clearly identify the limitations
of the effective two band $e_g$ TB description of manganite
systems. Our TB analysis was successful because in LaMnO$_3$ within
LSDA the $e_g$ bands are well-separated from the oxygen 2$p$ bands,
and neither the JT distortion nor the magnetic order change this
energy spacing, and thus the value of the effective hopping,
significantly. The dependence of the effective hopping parameters on
the energetic distance between the Mn $e_g$ and the O 2$p$ states (see
Eq.~(\ref{t_eff})) is visible as a 15-20\,\% difference between the
majority-spin and the minority-spin bandwidths and dispersion in the
ferromagnetic LSDA calculation for the cubic structure (see
Sec.~\ref{sec:cubic}), and also in the high-lying bands of the
antiferromagnetic LSDA band-structure shown in
Fig.~\ref{cubAFM}. However, in cases where the $e_g$ and O 2$p$ bands
overlap in energy, such as for example in our LSDA+$U$ calculation for
the purely JT distorted structure shown in Fig.~\ref{ldau}, the
effective $e_g$ TB analysis fails, and the O 2$p$ levels have to be
taken into account explicitly.

The energy shift of the occupied $d$ states relative to the oxygen $p$
states within the LSDA+$U$ method is mainly caused by the so-called
``double-counting correction'', which attempts to correct for those
contributions of the electron-electron interaction that are accounted
for in both the LSDA and the local Hartree-Fock (``+$U$'')
treatment. Since the double-counting correction is notoriously
ill-defined, this raises the question of whether such level shifts due
to the electron-electron interaction and the resulting substantial
renormalization of the effective hopping parameters are real effects,
or whether this is an artifact of the LSDA+$U$ scheme, which only
accounts for the static (mean-field) electron-electron interaction
between the transition metal $d$ states, while leaving the O 2$p$
states unchanged (we point out that the same problem is also present
within an LDA+DMFT treatment of electronic correlations). Optical
evidence (see Ref.~\onlinecite{Quijada_et_al:1998}) suggests that the
O 2$p$ bands in manganites are located about 4\,eV below the Fermi
level, consistent with the LSDA result, but more detailed
investigations of the energy separation between the Mn $e_g$ and O
2$p$ bands will be useful for future studies.


\begin{acknowledgments}
  This work was supported by the MRSEC Program of the National Science
  Foundation under award number DMR-0213574 (C.E.) and by the
  Department of Energy under grant number ER-46169 (A.J.M. and C.L.).
\end{acknowledgments}

\bibliography{../../literature,footnotes.bib}

\begin{thebibliography}{34}
\bibitem[{foo({\natexlab{a}})}]{footnote}
\bibinfo{note}{We use the notation of Ref.~\onlinecite{Bradley/Cracknell:Book}
  to denote the high-symmetry points within cubic, tetragonal, and orthorhombic
  Brillouin zones}.
\bibitem[{foo({\natexlab{b}})}]{footnote:notation}
\bibinfo{note}{Note that different conventions for the JT coupling are used
  throughout the literature, so that care must be taken when comparing values
  from different sources}.
\end{thebibliography}

\end{document}